\documentclass[12pt]{iopart}
\usepackage{iopams} 
\usepackage{graphicx}
\usepackage{xcolor}
\usepackage{enumitem}
\usepackage{cleveref}
\usepackage[numbers]{natbib}
\usepackage{indentfirst}
\usepackage{stackrel}

\newcommand{\teta}{\tau_{\eta}}
\newcommand{\eqref}[1]{Eq.~(\ref{#1})} 
\newcommand{\parref}[1]{(Eq.~\ref{#1})}
\definecolor{alizarin}{rgb}{0.82, 0.1, 0.26}
\definecolor{amber}{rgb}{1.0, 0.49, 0.0}
\definecolor{battleshipgrey}{rgb}{0.52, 0.52, 0.51}

\begin{document}

\title{Shot noise multifractal model for turbulent pseudo-dissipation}

\author{Gabriel B. Apolin\'{a}rio$^1$ and Luca Moriconi$^1$}

\address{$^1$Instituto de F\'{i}sica, Universidade Federal do Rio de Janeiro,
Av. Athos da Silveira Ramos, 149, Centro de Tecnologia, Bloco A,
Cidade Universit\'{a}ria, Rio de Janeiro, RJ, 21941-972}
\ead{gapolinario@pos.if.ufrj.br}
\vspace{10pt}
\begin{indented}
\item[]March 2020
\end{indented}

\begin{abstract}
Multiplicative cascades have been used in turbulence to generate fields with multifractal statistics and long-range correlations.
Examples of continuous and causal stochastic processes which generate such a random field have been carefully discussed in the literature. Here a causal lognormal stochastic process is built to represent the dynamics of pseudo-dissipation in a Lagrangian trajectory. It is introduced as the solution of a stochastic differential equation, driven by a source of noise which has sudden jumps at periodic intervals, its period being the dissipative time scale of the flow. This random field has scale invariance for a continuum of scales, and displays discontinuous jumps in time, with a smooth time evolution below the Kolmogorov scale. Its multifractal and correlation properties are demonstrated numerically.
\end{abstract}

\section{Introduction}

The first observations of intermittency in turbulence were reported quite a long time ago \citep{batchelor1949} and were first addressed theoretically in the work of Kolmogorov and Obukhov \citep{kolmogorov1962,obukhov1962}. In these works, scale dependent observables were postulated as the relevant quantities in the study of fluctuations in the turbulent inertial range.
The theory also hypothesized that the kinetic energy dissipation, a positive quantity, follows a lognormal probability distribution. This observation is remarkably accurate, as reported in experiments and numerical simulations \citep{yeungpope1989,dubrulle2019}.
Furthermore, experimental measurements of the kinetic energy dissipation
revealed long-range power-law correlations \citep{gurvitch1963,pond1965}, another key feature of turbulent fields.
Multifractal random fields have been a tool to describe and understand turbulent fields with such statistical properties, but their derivation on a first-principle basis is still an open problem.

The origin of the statistical distribution of the kinetic energy dissipation has been connected to the Richardson energy cascade through several phenomenological models, beginning with discrete cascade models \citep{novikov1964,yaglom1966,frisch1978,schertzer1984,benzi1984,meneveau1991,vainshtein1994,frisch1995}.
These models describe the distribution of energy dissipation across length scales in a turbulent field, from the large energy-injection scale, down to the much smaller dissipation length scales.
The energy transfer process is inviscid, as proposed in the hypotheses of the Kolmogorov 1941 theory \citep{kolmogorov1941dissipation,frisch1995}, and
dissipation only happens at the smallest relevant scales, close to the Kolmogorov length.
At the largest scale, $L$, the amount of energy transferred per unit time to the smaller scales is equal to the amount injected by the external force. This rate of energy transference is called $\varepsilon_0$. The next smaller length scale considered is $\ell_1 = L/\lambda$, where $\lambda > 1$ is the scale ratio constant specified by each model. The energy injection rate is broken down into a number of smaller shares at the smaller scale $\ell_1$, and this process carries the energy dissipation cascade forward. Each share of the energy dissipation is often called an \textit{eddy}, to highlight the geometric aspect of the energy cascade.
If each eddy is broken down into $N$ smaller eddies,
through each of the smaller eddies there is a flow of energy of value $W_1 \varepsilon_0 / N$. $W_1$ is a random variable, and the only requirements on it are that it is positive and with a mean value of one.
This demand on the variable $W_1$ guarantees a statistically inviscid cascade, meaning that the total energy flux through any length scale is always $\varepsilon_0$ on average. Repeated multiple times until the energy reaches the dissipation length, $\eta = L / \lambda^n$, this process generates an energy cascade, where $n$ is the total number of steps in it.
Therefore, the energy dissipation rate through the scale $\eta$ is a random variable, given by
\begin{equation} \label{eq:discrete-cascade}
	\varepsilon_n = W_1 W_2 \cdots W_n \ \varepsilon_0 \ .
\end{equation}
If the $W_i$ factors in this model are equally and independently distributed, the probability distribution for the small-scale energy dissipation is well approximated by a lognormal in the limit $n \to \infty$, as guaranteed by the central limit theorem.

This is a simple way to elicit the relevance of the lognormal distribution and its connection to the energy cascade. Different discrete models rely on this basis, with varying proposals for the way the energy is split at each step, and for the probability distribution of the $W_i$ factors.
Nevertheless, 
the central limit theorem does not apply in cases where $W_i$ display large fluctuations or strong correlations. In these cases, deviations from the lognormal distribution are significant and general concepts from large deviation theory \citep{touchette2009} should be applied instead of the central limit theorem, with the lognormal being only a quadratic approximation to the general result. Exact results for correlated factors can be found in some cases, as those indicated in Ref. \citep{cluselbertin2008}.
Furthermore,
lognormal fluctuations cannot be the precisely correct distribution of the energy dissipation, as discussed in Refs. \citep{frisch1995,paladin1987anomalous}, since the corresponding structure function exponents violate Carleman's criterion \citep{carleman1922}, a general requirement on the moments of a probability distribution.
These violations are only manifest for high order moments, though, and the lognormal is still a valid approximation at low orders.

The cascade models described by \eqref{eq:discrete-cascade} had the limitation of being discrete and of possessing a special scale ratio between neighboring scales, customarily $\lambda=2$. It was noticed early \citep{mandelbrot1972} that this special scale ratio should not be present, since turbulent energy dissipation displays multifractal properties for any chosen scale. Instead, a description in which arbitrary values of $\lambda$ are valid and produces multifractal statistics should be preferred and investigated, as pointed out by Mandelbrot in Ref.~\citep{mandelbrot1972}.
Furthermore, the discrete models have been able to account for scale-locality of the energy transfer process, but did not account for time and space correlations. In other words, such fields had no causal structure and could not be connected to the Navier-Stokes equations.
These issues
were addressed in several works \citep{biferale1998mimicking,boffetta1999pair,schmittmarsan2001,barralmandelbrot2002,muzy2002,pereira2018} which build sequential and multifractal stochastic processes.

This work describes a causal stochastic process driven by discrete and periodic random jumps, which is used to model the dynamical and multifractal properties of Lagrangian pseudo-dissipation. Discrete noise provides a model dynamics which is regular at scales below the Kolmogorov length scale, conforming to the Richardson cascade view of turbulence.
The dynamics we observe, on large scales, exhibits multifractal statistics and long-range correlations similarly to models driven by continuous noise, demonstrating the possibility to apply discrete (shot) noise in effective models of turbulence.
Our main motivation
here relies on the fact that the time evolution of local Lagrangian observables is sensitive to the existence of spacetime localized perturbations of the turbulent flow, such as vortex tubes. This can be particularly appreciated, for instance, in the dynamics of spheroids in turbulent flows, which depends on small scale properties of the velocity gradient tensor \citep{parsa2012,voth2017}. Their tumbling is marked by regular evolution disrupted by intense jumps, indicating that a modeling based in terms of shot noise sources might explain their behavior. The main characteristics of a turbulent flow which lead to the preferential alignment of these spheroids is still a problem under wide investigation, with possible applications in industrial and natural flows.

It is compelling to note that the analytical advantages of the lognormal formulation make it suitable for applications in several other fields where intermittent fluctuations play a role, besides turbulence, such as in financial economics \citep{mandelbrot1997,ghashghaie1996,liu1999}, cosmology \citep{coles1991} and condensed matter systems \citep{kravtsov1997,serbyn2017}.
Furthermore, the construction of a causal equation for a multifractal field driven by shot noise is far from trivial, requiring the use of a general version of It{\^o'}s lemma, including the contributions from discontinuities \citep{protter2005,klebaner2012}. This lemma and its application to the random field in case are discussed in detail in the following discussions.

Focusing on turbulence, it turns out, from experimental evidence \citep{yeungpope1989,dubrulle2019}, that several positive-definite observables like the kinetic energy dissipation, kinetic energy pseudo-dissipation, enstrophy, and the absolute value of acceleration can be reasonably well described by lognormal distributions, with a particularly good accuracy being achieved for pseudo-dissipation.
In the mentioned work, it is also remarked that the statistical moments of dissipation and enstrophy seem to approach those of the lognormal distributions as the Reynolds number increases. Yet,
since the lognormal can only be a good approximation, but not a complete solution, further are required to settle this issue.

This paper is organized as follows: In Section~\ref{sec:stat-dissip}, we discuss previous theoretical models and results about the general statistics of positive-definite scalar quantities of interest in turbulence. Next, in Section~\ref{sec:stoc-lag-dissip}, we address stochastic models applied to the evolution and statistics of Lagrangian pseudo-dissipation, including the statistical properties known from the previous section, and a description of the non-Markovian shot noise process which is the main object of this work. Then, in Section~\ref{sec:numerical}, the numerical procedure to obtain an ensemble of solutions of the proposed stochastic process is described, followed by the results obtained from the simulations, presented in Section~\ref{sec:results}. Concluding remarks are detailed in Section~\ref{sec:conclusion}, along with possible extensions, further questions and applications.

\section{Statistics of Turbulent Energy Dissipation and Pseudo-Dissipation} \label{sec:stat-dissip}

The first theoretical results in the statistical theory of turbulence \citep{kolmogorov1941dissipation,frisch1995}, established the picture of the turbulent cascade on a mathematical ground. This early description of Kolmogorov regards the mean behavior of the inertial range statistics of turbulent velocity fields, but not its fluctuations. Later studies of turbulent fluctuations, leading to the multifractal picture,
revealed that the K41 velocity field is an exactly self-similar field of Hurst exponent $1/3$, that is, a monofractal. This field is homogeneous in space, in contrast to the complex and concentrated structures which form in isotropic flows, revealed by direct numerical simulations and experiments \citep{ishihara2009,debue2018prf,dubrulle2019}.

Multifractal fields have been proposed as general models to the  turbulent velocity field in the inertial range, although it remains an open problem to fully characterize this multifractal field and its statistical properties.
For the purpose of modeling a positive-definite scalar field, consider a generic $d$-dimensional multifractal random field $\varepsilon_{\eta}$, which may depend on the spatial variable $\boldsymbol{x}$ and on time $t$, and with a dissipative length scale $\eta$.
The basic statistical properties of this random field are compatible with the features of the discrete cascade models and with experimental and numerical realizations of several observables in turbulence.
The statistical
moments $\left\langle(\varepsilon_{\eta})^{q}\right\rangle$ of this field can be calculated as an ensemble average or as a self-average over a single time series, assuming ergodicity, which has been numerically verified in turbulent fields \citep{galanti2004,djenidi2013ergodic}. These moments satisfy the relation
\begin{equation} \label{eq:bare-moments}
\left\langle(\varepsilon_{\eta})^{q}\right\rangle = A(q) \ \eta^{K(q)} \ 
\end{equation}
in the limit of $\eta \to 0$ (equivalent to $\mathrm{Re} \to \infty$).
In this equation,
$A(q)$ is a $q$-dependent constant and $K(q)$ is a characteristic function of the multifractal field, connected to how structures at different scales spread across space.
In particular, a lognormal distribution for $\varepsilon_{\eta}$ corresponds to $K(q) = \mu q (q-1) / 2$, where $\mu$ is called the intermittency parameter, which measures the intensity of the fluctuations of this field. In the case of Eulerian three-dimensional turbulence, $\mu=0.2$ \citep{stolovitzky1994,praskovsky1997,chen1997}.
A monofractal field,
in its turn, would have $K(q) = \mu q$.

The variable $\varepsilon_{\eta}$ is a bare field, since it is defined at the dissipative scale. The multifractal hypothesis makes predictions for the behavior of coarse grainings of $\varepsilon_{\eta}$ as well, which are defined as local averages of the original field at the scale $\eta' > \eta$:
\begin{equation} \label{eq:dissip-coarse-ddim}
	\varepsilon_{\eta'}(\boldsymbol{x},t) = \frac{1}{V_d(\eta')} \int_{\mathcal{B}_{\eta'}(\boldsymbol{x})} \varepsilon_{\eta}(\boldsymbol{x+r},t) d\boldsymbol{r} \ ,
\end{equation}
where $\mathcal{B}_{\eta'}(\boldsymbol{x})$ is a $d$-dimensional ball of radius $\eta'$ and center $\boldsymbol{x}$, with its $d$-dimensional volume indicated by $V_d(\eta')$.
In particular, the statistical moments of a coarse-grained multifractal field obey the same statistical behavior as the bare field,
\begin{equation} \label{eq:dressed-moments}
\left\langle(\varepsilon_{\eta'})^{q}\right\rangle = A'(q) \ \eta'^{K(q)} \ ,
\end{equation}
at scales larger than the bare scale $\eta$ and up to some critical moment $q_{\mathrm{crit}}$, beyond which this scaling becomes linear \citep{schmitt1994,lashermes2004}.
It is vital to know these properties for coarse-grained fields for two main reasons. First, a coarse-grained field is all experimentalists have access to. And second, the features of coarse-grained fields are a fundamental ingredient in Kolmogorov's refined similarity hypothesis, according to which the inertial range statistical properties at scale $\ell$ depend only on $\ell$ itself and on the kinetic energy dissipation coarse-grained at this scale, $\varepsilon_{\ell}$. Thus, the verification that a given set of data does display multifractal statistics compels to the study of its coarse-grained properties.

For the general field $\varepsilon_{\eta}$, given that $\gamma_{\eta} \equiv \ln \varepsilon_{\eta}$, its autocorrelation function decays logarithmically with the distance between the points,
\begin{equation} \label{eq:ln-correlation}
\langle \gamma_{\eta}(0) \gamma_{\eta}(\boldsymbol{r}) \rangle = C - \frac{\sigma^2}{\ln \lambda} \ln |\boldsymbol{r}| \ .
\end{equation}
This property can be easily verified for the the discrete cascade models \citep{arneodo1998prl,arneodo1998jmathphys,schmitt2003}, in which case $\sigma^2 = \mathrm{Var}[ \ln W ]$ and $C = \langle \ln W \rangle^2 n^2 + \sigma^2 n$, where $n$ is the depth of the cascade and $\lambda$ the scale ratio of the model.
The Fourier transform of this expression corresponds to the ubiquitous $1/f$ power spectrum,
\begin{equation}
E_{\eta}(k) \approx k^{-1} \ .
\end{equation}
This is a common feature of intermittent fields in general \citep{schertzer1987,schertzer1991,bacry2001} and is also valid for coarse-grained fields.
The properties just presented, (\ref{eq:bare-moments}), (\ref{eq:dressed-moments}), and (\ref{eq:ln-correlation}) are the main characteristics of a multifractal field.

The need for random fields with such properties has been a development of the work of Kolmogorov and Obukhov \citep{kolmogorov1962}.  To take fluctuations into account, it was postulated in this work that the kinetic energy dissipation field follows a lognormal distribution with
\begin{equation} \label{eq:kolmogorov-var}
	\mathrm{Var}[\ln \varepsilon_{\ell}] = - \mu \ln (\ell/L) + C \ ,
\end{equation}
where $L$ is the integral length scale, $\ell$ is the observation scale in the inertial range, $\eta \ll \ell \ll L$, and C is an arbitrary constant. The intermittency parameter $\mu$, the same as in the expression for $K(q)$, was historically introduced in this expression. Mandelbrot \citep{mandelbrot1972} noticed that a random field built as the exponential of a Gaussian field, $\varepsilon_{\eta} \propto \mathrm{exp}\{\sqrt{\mu} X\}$, would satisfy these properties. His construction was mathematically formalized in \citep{kahane1985,robert2010}, and the modern understanding of such random fields has led to explicit frameworks in the Eulerian \citep{pereira2016} and Lagrangian context \citep{pereira2018}, which approximate the known statistics of turbulent fields.
It was later realized \citep{yaglom1966} that the intermittency parameter is also responsible for the power-law correlations of the kinetic energy dissipation, in the form
\begin{equation} \label{eq:gurvitch-corr}
\langle \varepsilon_{\eta}(\boldsymbol{r}) \varepsilon_{\eta}(\boldsymbol{r+\delta r}) \rangle \propto (L/\ell)^{\mu} \ ,
\end{equation}
where $\ell \equiv |\boldsymbol{\delta r}|$ and the parameter $\mu$ is apparent as well.
The cascade models were built to explain these statistical features.

The specific random fields considered in this work, as well as \citep{schmitt2003,perpete2011,pereira2018} are one-dimensional and depend only on time, since they correspond to some positive-definite observable following a Lagrangian trajectory of the flow.
The Lagrangian view is connected to the space-time structure of the energy dissipation cascade, since eddies are carried by the flow, their statistical distribution is somehow influenced by the transport properties of the turbulent velocity field, consequently leading to the cascade process.
And Lagrangian observables such as velocity differences and velocity gradients have been argued to display scaling and intermittent behavior, following a Lagrangian refined similarity hypothesis, in an equivalent manner to the Eulerian framework. Lagrangian velocity difference structure functions, for instance, are believed to scale as
\begin{equation} \label{eq:lagrangian-exponents}
	\langle \big( \delta u_i(\tau) \big)^n \rangle \propto (\langle \varepsilon_{\tau} \rangle \tau)^{\xi_n}
\end{equation}
in the Lagrangian inertial range, $\teta \ll \tau \ll T$, between the dissipative time scale $\teta$ and the integral time scale $T$.
The coarse-grained Lagrangian kinetic energy dissipation, $\varepsilon_{\tau}$, is defined in terms of its bare counterpart, $\varepsilon_{\teta}$, in analogy with \eqref{eq:dissip-coarse-ddim}:
\begin{equation} \label{eq:dissip-coarse-1dim}
	\varepsilon_{\tau}(t) = \frac{1}{\tau} \int_{t}^{t+\tau} dt'
	\varepsilon_{\teta}(t') \ .
\end{equation}
Its average value $\langle \varepsilon_{\tau} \rangle$ is constant due to the stationarity of the turbulent flow. The dissipative time scale is determined from dimensional analysis as the Lagrangian analogue of the dissipative length scale, and is defined as $\teta = \eta^{2/3} \varepsilon_0^{-1/3}$ and the Lagrangian integral time is defined in terms of the velocity two-point autocorrelation $\rho_L(\tau)$, as $T = \int_0^{\infty} \rho_L(\tau) d\tau$.
In the K41 framework, the scaling exponents of velocity difference structure functions grow linearly, and the equivalent relation in the Lagrangian view is that $\xi_n = n/2$. This can be understood with the work of Borgas \citep{borgas1993}, which connects Lagrangian and Eulerian self-similarity.
\eqref{eq:lagrangian-exponents} has been numerically verified in \citep{benzi2009,sawfordyeung2011,barjona2017}, and it is notable that finite Reynolds effects are more pronounced in the Lagrangian frame, making measurements even more difficult \citep{yeung2002}.

The discrete cascades display the same statistical properties as the small scale multifractal field proposed by Mandelbrot \citep{mandelbrot1972}, yet for a special scale ratio. This was one of the main critiques of Mandelbrot. Continuous multiplicative cascades have been investigated since then, with the objective of building models with continuous scale invariance, which the discrete models explicitly broke. The continuous model of Mandelbrot is a direct extension of the discrete models, in which the energy dissipation at each scale is a continuous product of random factors, with energy dissipation being only statistically conserved along the cascade. That is a straightforward, albeit nonrigorous, translation of \eqref{eq:discrete-cascade} to the continuum. In the discrete case, the central limit theorem ensures that the energy dissipation follows an essentially lognormal probability distribution, if there are enough steps in the discrete cascade
and its
factors are independent.
In the continuous case this conclusion holds as well, this result was conjectured in \citep{mandelbrot1972} and proven in \citep{kahane1985}, a work which gave solid mathematical foundations to the continuous cascade approach and elicited its statistical properties. Since \citep{kahane1985}, this continuous stochastic process with multifractal statistics is called Gaussian Multiplicative Chaos, in connection with the standard additive chaos (more commonly called the Wiener process) \citep{rhodes2014,berestycki2017}.
Continuous cascade models have inspired a huge body of work to this day, both in turbulence and in other areas of research such as quantitative finance \citep{duchon2012}, quantum gravity in two dimensions \citep{duplantier2014} and random matrix theory \citep{fyodorov2014}.

Another critique of Mandelbrot on the discrete cascades was the absence of a space-time causal structure. The only causal connection in these models is between length scales, a relation which cannot be easily translated to a space-time distribution of turbulent structures or of energy dissipation.
The pursuit
of effective stochastic models in turbulence dates back to Refs. \citep{thomson1987,girimaji1990diffusion}, with models for the velocity and velocity gradient, respectively.
Sequential stochastic models for
multifractal fields were then proposed in \citep{biferale1998mimicking,boffetta1999pair,schmitt2003}.

In Ref.~\citep{schmitt2003}, 
analytical expressions for the statistical moments and two-point correlation functions of a multifractal stochastic process are proved, in agreement with the multifractal hypothesis and providing a continuous-in-scale extension of the discrete cascade models.
This stochastic process, though, does not generate a stationary state solution, an issue which was resolved in \citep{pereira2018}.
The stochastic process of Ref. \citep{schmitt2003} for the evolution of the field $\varepsilon_S$ is
\begin{equation} \label{eq:schmitt-continuous}
\eqalign{
	&\mathrm{d} \varepsilon_S(t)=\sqrt{\mu} \ \varepsilon_S(t)\left( \frac{1}{\sqrt{\teta}} d W(t)+\frac{1}{2}(1-\beta_S(t)) d t\right) \ , \\
	&\beta_S(t)=\teta \sqrt{\mu} \int_{t+\teta-T}^{t}(t+\teta-u)^{-3 / 2} d W(u) \ .
}
\end{equation}
In this equation, $T$ is the integral time scale, $\teta$ the Kolmogorov dissipative time scale, $\mu$ the intermittency parameter (as defined by Kolmogorov, \eqref{eq:kolmogorov-var}) and $W(t)$ is a standard Wiener process.
The
$\beta_S(t)$ term illustrates the important role that non-Markovian correlations perform in a multifractal time series and was introduced as an analogue in time of the cascade happening in scale space, as described in Ref.~\citep{schmitt2003}.
This contribution, driven by the same random noise $W(t)$ as the main equation, is responsible for the long-range correlation of $\varepsilon_S(t)$.

Furthermore, \eqref{eq:schmitt-continuous} can be viewed as the exponential of a Gaussian process, in the way first proposed by Mandelbrot for multifractal fields.
With It\^{o}'s lemma, the underlying process is found to be
\begin{equation} \label{eq:schmitt-x}
	d X_S(t)=\sqrt{\frac{\mu}{\teta}} \ d W(t) - \frac{1}{2} \beta_S(t) d t \ ,
\end{equation}
where $\varepsilon_S(t) = \exp X_S(t)$.
This perspective is a handy connection with the product of random factors of \eqref{eq:discrete-cascade}, where the discrete product has been replaced by a continuous sum over Gaussian random variables.

Nevertheless, \eqref{eq:schmitt-continuous} cannot be used to accurately describe turbulent fields for it does not generate a stationary state. This point was addressed in Ref. \citep{pereira2018}, in which the causal framework which evolves to multifractal stationary states was introduced. It is applied to the description of the Lagrangian pseudo-dissipation, $\varphi = \sum_{i,j=1,2,3}(\partial_i u_j)^2$, and Lagrangian velocity-gradients, and is explicitly given by the following stochastic differential equation:
\begin{equation} \label{eq:pereira-x}
\eqalign{
	d X_P(t)&=\left[-\frac{1}{T} X_P(t)+\beta_P(t)\right] d t+\frac{1}{\sqrt{\tau_{\eta}}} dW(t) \ , \\
	\beta_P(t)&=-\frac{1}{2} \int_{s=-\infty}^{t} \frac{1}{\left(t-s+\tau_{\eta}\right)^{3 / 2}} dW(s) \ ,
}
\end{equation}
where the pseudo-dissipation $\varphi_P(t)$ is given by an exponential of the underlying $X_P(t)$ process, explicitly:
\begin{equation}
	\varphi_P(t)=\frac{1}{\tau_{\eta}^{2}} \exp \left\{ \sqrt{\mu} X_P(t)-\frac{\mu_{l}}{2} \mathbb{E}\left[X_P^{2}\right] \right\} \ .
\end{equation}
In this work, the statistical moments and autocorrelation of $X_P(t)$ and $\varphi_P(t)$ were shown to follow multifractal laws, in the limit of $\teta \to 0$ (corresponding to infinite Reynolds number).

An extension
of Eqs.~(\ref{eq:schmitt-continuous}) and (\ref{eq:pereira-x}) and a connection to fractional Brownian motion was made in \citep{chevillard2017}, where it is shown that the non-Markovian contribution ($\beta$) pertains to a family of noisy integrators indexed by the Hurst exponent $H \in [0,1]$. The exponent $3/2$ corresponds to $H=0$, the roughest instance, while positive $H$ represents more regular stochastic fields, with an exponent $3/2-H$ in the $\beta$ term.
The term
between parentheses in the equation for $\varepsilon_S$ \parref{eq:schmitt-x} can be viewed as generating a fractional Brownian Motion of Hurst exponent $H=0$, as explained in Ref.~\citep{chevillard2017}. The addition of linear damping in \eqref{eq:pereira-x} is responsible for the change from a fractional Brownian Motion to a fractional Ornstein-Uhlenbeck process, producing a stationary process for the pseudo-dissipation.

The model
of \eqref{eq:pereira-x} also leads to a successful stochastic velocity gradient model \citep{pereira2018} which extends the Recent Fluid Deformation model \citep{chevillard2006} to high Reynolds numbers. It provides a reasonable reproduction of the orientation statistics of rod-like spheroid tracers in turbulent flows, but not disc-like objects, both described by Jeffery's equation \citep{parsa2012,jeffery1922}.

It is noted
that this formulation based on the exponential of a Gaussian process is essentially multifractal and focuses on lognormal statistics.
The extension of $X$ from a normal to a stable distribution, which would bridge a connection with more general stochastic models, was addressed in the discrete model of Ref.~\citep{perpete2011}.
Some discrete cascades generate simple fractals (monofractals) in the Eulerian context, such as the random-$\beta$ model \citep{benzi1984}, while in the formalism of Eqs.~(\ref{eq:schmitt-continuous}) and (\ref{eq:pereira-x}), only the trivial limit $\mu \to 0$ would generate a static dissipation/pseudo-dissipation, which corresponds to the eddy-filling interpretation of the K41 framework.

\section{Stochastic Models of Lagrangian Pseudo-Dissipation} \label{sec:stoc-lag-dissip}

An alternative formulation of multiplicative chaos was done in Ref.~\citep{perpete2011}, where a time-discretized causal multifractal process was introduced.
In this work it was proven that a process which is discrete in time may have a continuous scale ratio, along with the statistical features of multifractality, in the limit of large integral time (corresponding to infinite Reynolds number).

The stochastic process of \citep{perpete2011}, with a dissipative timescale $\teta$ and a large timescale $T = N \teta$, is described by
\begin{equation} \label{eq:perpete-x}
    X_D(t) = \frac{1}{\sqrt{\teta}} \sum_{k = 0}^{N-1} (k+1)^{-1/2} \alpha_{t-k} \ ,
\end{equation}
where $\alpha_k$ are independent and identically distributed Gaussian random variables of zero mean and standard deviation $\sqrt{\teta}$. The time, unlike in the previous examples, is only defined for integer $t$. This process also possesses long-term memory over the integral time scale, in connection with the $\beta$ term in Eqs.~(\ref{eq:schmitt-continuous}) and (\ref{eq:pereira-x}).
The multifractal process corresponding to \eqref{eq:perpete-x} is likewise given by its exponential,
\begin{equation}
    \varphi_D(t) = \varphi_0 \exp \big( \sqrt{\mu} X_D(t) - \mu \mathbb{E}[X_D^2]/2 \big) \ \mbox{,}
\end{equation}
which has lognormal and long-range correlated statistics.
\eqref{eq:perpete-x} reaches a stationary state with the following multifractal properties:
\begin{enumerate}[label=(\roman*)]
    \item \label{it:bare}
    Its moments satisfy
    \begin{equation} \label{eq:perpete-moments}
    \mathbb{E}[ \varphi_D^q ] =
    C_q \left( \frac{\teta}{T} \right)^{-K(q)} \ \mbox{,}
    \end{equation}
    with $K(q) = \mu \ q (q-1) / 2$, conforming to the lognormal statistics, and $C_q$ is a factor which can be
    exactly calculated.
    \item \label{it:dress}
    The coarse-grained moments of $\varphi_D$ satisfy a similar relation with the same exponents:
    \begin{equation} \label{eq:perpete-moment}
    \mathbb{E}[ (\varphi_D)_{\tau}^q ] \approx
    c_q \left( \frac{\tau}{T} \right)^{-K(q)} \ \mbox{,}
    \end{equation}
    where the coarse-grained field is defined as a moving average with a window of $\tau = m \teta$:
    \begin{equation}
    (\varphi_D)_{\tau}(t) = \frac{1}{\tau} \sum_{k=t}^{t+m-1} \varphi_D(k) \ .
    \end{equation}
    Relation (\ref{eq:perpete-moment}) is asymptotic, and is valid in the limit of $T$ going to infinity, with the ratio $\tau/T$ kept fixed. Furthermore, $q$ must be such that $K(q) < q-1$.
    The existence of upper and lower bounds on $c_q$ was demonstrated in \citep{perpete2011}, while precise values would have to be inspected numerically.
    \item The autocovariance of this process, in the same limit of $T \to \infty$,
    converges to
    \begin{equation} \label{eq:perpete-cov}
    \mathrm{Cov}[\varphi_D(t), \varphi_D(t+\tau)] \approx
    - \mu \ln(\tau/T) \ \mbox{.}
    \end{equation}
\end{enumerate}

Having Eqs.~(\ref{eq:perpete-moments}-\ref{eq:perpete-cov}) in mind, we are going to construct a stochastic differential equation for a multifractal process which inherits features from the continuous and discrete instances just described. This stochastic field takes into account the small scales in a dynamic manner, such that it follows a smooth time evolution on scales below the Kolmogorov time, but still shows roughness and multifractal behavior on timescales much larger than that. This picture is inspired by the Kolmogorov phenomenology, in which dissipation can be neglected in the inertial range, while it acts in smoothing out the velocity field in the dissipative scale.

Furthermore, the refined similarity hypothesis of Kolmogorov \citep{kolmogorov1962,frisch1995} states that, in the limit of infinite Reynolds numbers, all the statistical properties at scale $\ell$ are uniquely and universally determined by the scale itself and the mean energy dissipation rate coarse-grained at this scale, $\varepsilon_{\ell}$.
By this hypothesis, it is expected that a variety of noise sources generate similar behavior, due to an independence of the inertial range properties on the details of the dissipative dynamics.
For this reason, several large scale observables of the random field stirred by discrete noise should converge to the same quantities as fields driven by Wiener noise.

The stochastic process we describe in this work is a model for Lagrangian pseudo-dissipation forced by a discrete noise source with a long time memory.
This stochastic process evolves in continuous time, while being driven by a random force which is periodic in time and only acts in discrete instants. A stationary state arises as solution of this process and its statistical properties are investigated, in comparison with the properties of the multifractal random fields already described in the literature.
Shot noise is used
with the aim of modeling quiescent regions of Lagrangian fields: It is observed that correlated events of intense fluctuations are interspersed by regions of damped statistical fluctuations. These intervals of rest can be clearly observed in time series of tumbling spheroids following Lagrangian trajectories \citep{parsa2012}.
Studies of discrete noise (often called shot noise) or a mixture of discrete and continuous noise (or jump-diffusion) have been pursued in others areas of knowledge as well, such as financial economics \citep{cox1976,merton1976,ball1993,duffie2000,das2002}, neuronal systems \citep{giraudo1997,patel2008,sacerdote2013}, atomic physics \citep{chudley1961,ellis1993,funke1993,montalenti1999}, biomedicine \citep{grenander1994} and image recognition \citep{srivastava2002}.

The presence of
non-Markovian noise in Eqs.~(\ref{eq:schmitt-continuous}), (\ref{eq:pereira-x}) and (\ref{eq:perpete-x}), is connected to the observed long-range correlations of the pseudo-dissipation field.
It is remarked in Ref.~\citep{chevillard2017} the importance that correlations between the external noise $W(t)$ and the drift contribution $\beta(t)$ have in generating multifractal fields.
Non-Markovian noise is also a consequence of the effective nature of one-dimensional Lagrangian models: The spatial degrees of freedom of Eulerian models are integrated out, and as a result complex long-range memory effects arise in Lagrangian models. This is in consonance with renormalization group studies of effective stochastic models: Stochastic equations defined at a microscopic length scale and driven by white noise, when coarse-grained, display non-trivial memory effects in their new external noise term \citep{goldenfeld1998,hou2001}.

Explicitly, we consider the stochastic process given by the stationary solution of the following differential equation:
\begin{equation} \label{eq:jump-scalar}
    dX(t) = \left( -\frac{1}{T} X(t^-) + \beta(t) \right) dt + \frac{1}{\sqrt{\teta}} \sum_{\teta \ell \leq t} \alpha_{\ell} \ \delta(t-\teta \ell) \ dt \ .
\end{equation}
The first term on the RHS corresponds to a drift in a usual Langevin equation, and has the same form as the drift term in \eqref{eq:pereira-x}. The first contribution in this term is responsible for correlations of the $X(t)$ random field of characteristic time $T$, while the second is in charge of the multifractal correlations in the solution, with a similar role to the long-memory term present in Eqs. (\ref{eq:pereira-x}) and (\ref{eq:perpete-x}). The second term in \eqref{eq:jump-scalar} accounts for the discrete random jump contributions. These jumps occur at periodic intervals of length $\teta$ and have an intensity $\alpha_{\ell}$, which is a Gaussian random variable of zero mean and standard deviation of $\sqrt{\teta}$. Each value of $\ell$ corresponds to a jump instant $\ell\teta$, hence the sum is carried for all jump times prior to the observation time $t$.

It is also important to observe the presence of the $t^-$ in the first term which represents an instant infinitesimally preceding the current observation instant. In a stochastic process with jumps, this kind of care is needed, because the current state of the system (at $t$) depends on the continuous evolution up to time $t^-$ and on the value of a jump which may have happened exactly at the instant $t$, and therefore does not affect the previous state of the system, only its future evolution. For this reason, the state $X(t^-)$ and a jump $\alpha_{\ell}$ happening exactly at $\ell \teta = t$ are completely independent events.
In the traditional notation of point processes \citep{protter2005,klebaner2012}, continuous random fields are taken to be \textit{c\`{a}dl\`{a}g}, a French acronym for \textit{continuous on the right and limit on the left}. This denomination means that jumps occur exactly at the instant $t_{\ell}$, while the left-limit at $t_{\ell}^-$, is not at all influenced by the jump term. Then, for a discontinuous random field $f(t)$ with a random jump happening at $t_{\ell}$, being \textit{c\`{a}dl\`{a}g} is equivalent to
\begin{equation}
	\lim_{t \to t_{\ell}^-} f(t) \neq f(t_{\ell}) \ \ \mbox{and} \ \
	\lim_{t \to t_{\ell}^+} f(t) = f(t_{\ell}) \ .
\end{equation}

The drift term in \eqref{eq:jump-scalar} contains a random contribution, $\beta(t)$, in correspondence with the long-term memory random contributions in \citep{schmitt2003,chevillard2017,pereira2018}, thus characterizing this model as a version of the Fractional Ornstein-Uhlenbeck process. The expression for this term is
\begin{equation} \label{eq:beta}
    \beta(t) = -\frac12 \sum_{\teta \ell \leq t}  \ \frac{\alpha_{\ell}}{(t-\teta \ell + \teta)^{3/2}} \ \mbox{,}
\end{equation}
where the $\alpha_{\ell}$ are exactly the same as those already sampled randomly for \eqref{eq:jump-scalar}.
The sum is also carried out over all jump times up to the time $t$.

The solution to this equation can be written explicitly in terms of a particular realization of the random jumps:
\begin{equation}
    \eqalign{
        X(t) &= \int_{s = t-T}^{t} \frac{e^{(s-t)/T}}{\sqrt{t-s+\teta}} \ \sum_{\ell} \alpha_{\ell} \ \delta(s-\teta \ell) \ ds \\
        &+ \frac{1}{\sqrt{T+\teta}} \int_{s=0}^t e^{(s-t)/T} \ \sum_{\ell} \alpha_{\ell} \ \delta(s-\teta \ell+T) \ ds \ ,
    }
\end{equation}
where, after integrating over the delta functions, we obtain
\begin{equation} \label{eq:x-solution}
\eqalign{
        X(t) &= \sum_{t-T < \teta \ell \leq t}
        \frac{e^{(\teta \ell-t)/T}}{\sqrt{t-\teta \ell+\teta}} \  \alpha_{\ell} \\
        &+ \frac{1}{\sqrt{T+\teta}} \sum_{0 < \teta \ell \leq t}
        e^{(\teta \ell-t-T)/T} \ \alpha_{\ell - T/\teta} \ .
}
\end{equation}
From this solution, several analytical properties of the stationary stochastic field can be calculated and compared to the results of numerical simulations and to the results of the continuous random field of \citep{pereira2018}.

Still, the solution in \eqref{eq:x-solution} has only Gaussian fluctuations. In analogy to what is done in the discrete \citep{perpete2011} and continuous settings \citep{schmitt2003,pereira2018}, the field with multifractal correlations is, in fact,
\begin{equation} \label{eq:exp-x}
    \varphi(t) = \varphi_0 \ \exp\{\sqrt{\mu} X(t)
    - \mu \mathbb{E}[X(t^-)^2]/2 \} \ ,
\end{equation}
where the mean value of this process is defined as $\varphi_0 = 1/\teta^2$, following the phenomenology of Kolmogorov \citep{girimaji1990diffusion}.
The variance of the $X(t)$ process, $\mathbb{E}[X^2(t)]$, can be calculated from the analytical solution, \eqref{eq:x-solution}:
\begin{equation} \label{eq:x2-variance}
\eqalign{
	\mathbb{E}[X^2(t)] &=
	\sum_{t-T \leq \teta\ell \leq t} \teta \frac{e^{2(\teta\ell-t)/T}}{t-\teta \ell + \teta}
	+ \frac{\teta}{T + \teta} \sum_{0 \leq \teta\ell \leq t} e^{2(\teta\ell-t)/T-2} \\
	&+ \frac{2 \teta}{\sqrt{T+\teta}} \sum_{t-T \leq \ell \leq t} \frac{e^{2(\teta\ell-t)/T-1}}{\sqrt{t-\teta \ell + \teta}} \ ,
}
\end{equation}
thus it can be seen simply as a function of time.

\eqref{eq:exp-x} also explains the choice of periodic discrete noise with period $\teta$, instead of the common choice of Poisson noise 
with an equal characteristic time, which is often what is referred to as shot noise \citep{morgado2016}. The variable $z = \exp{\sqrt{\mu}X}$, where $X$ is a sum of $N$ Gaussian random variables, follows a lognormal probability distribution. In the case of Poisson noise, the amplitudes of the jumps would be given by the normal distribution as well, but the number of jumps would be random, with a mean $N$, and $z$ would not follow a lognormal distribution exactly. In the limit of $N \to \infty$, though, both distributions coincide, by the central limit theorem.

Furthermore,
it is a consequence of multifractal theory that fluctuations of the dissipative scale exist, reaching below the Kolmogorov scale \citep{paladin1987anomalous,nelkin1990multifractal,frischvergassola1991,yakhot2006probability,arneodo2008universal}, whereas the smallest time scale in \eqref{eq:jump-scalar} is fixed at the Kolmogorov time, $\teta$.
This model with discrete jumps is effective at scales larger than the Kolmogorov one, but the smooth motion at sub-Kolmogorov scales is still an interesting feature connected to the role of viscosity in smoothing out fluctuations.

We performed, for the process described by the pair of Eqs. (\ref{eq:jump-scalar}) and (\ref{eq:exp-x}), numerical simulations to verify its statistical properties. The details of the numerical procedure are described in the next section.
We also
remark that an equation for the evolution of $\varphi(t)$ directly can be written with the use of It\^{o}'s lemma for semimartingale processes. This change of variables is useful to draw connections between the model for the pseudo-dissipation and other possible observables, such as the velocity or the velocity gradient.
The details of this procedure in the context of a stochastic equation with discontinuous jumps are described in Appendix A.

\section{Numerical Procedure} \label{sec:numerical}

Numerical solutions of the stochastic process described by \eqref{eq:jump-scalar} were calculated to verify the claims of its multifractal properties. The results are compared to the analytical and numerical results obtained in previous works, particularly \citep{schmitt2003,perpete2011,chevillard2017}.

The time evolution of \eqref{eq:jump-scalar} can be split in a deterministic contribution from the drift term, $(-X/T+\beta)$, and a jump term, proportional to a random jump intensity $\alpha_{\ell}$. There are sophisticated algorithms to obtain the solutions of general jump-diffusion equations, such as those illustrated in \citep{casella2011,gonccalves2014}, which provide a framework to deal with complex time-dependence in the drift or diffusion coefficients, cases where the solution cannot be obtained with a straightforward stochastic Euler algorithm. Instead, the diffusion term in \eqref{eq:jump-scalar} does not display any time-dependence, and the $\beta(t)$ term has a long-range memory, requiring a simpler algorithm in its implementation. With these considerations in mind, we have applied the Euler algorithm described in \citep{casella2011} for the simulation of the stochastic jump process of \eqref{eq:jump-scalar}.

The jumping times are known in advance, since they are periodic, and given simply by $(0,\teta,2 \teta,\ldots)$. For each interval between two jumps, $((\ell-1)\teta,\ell\teta)$, the drift term is simulated with an Euler algorithm, which is used to calculate $X(t_{\ell}^-)$. Then, the jump term is given by a random sampling and used to determine $X(t_{\ell})$.
To setup the initial condition for the simulation, jump intensities $\alpha_{\ell}$ are arbitrarily defined for a few complete integral times in the interval $[-T,0]$. The random jump intensities in this past interval are sampled exactly like the intensities in the core of the simulation, as Gaussian random variables of mean zero and standard deviation $\sqrt{\teta}$. The choice of $X(0)=0$ is made as well.
\eqref{eq:beta} depends on the whole time evolution of the system, hence a truncation in the past evolution is required. A complete integral time has been chosen since it provides accurate results in comparison with the theoretical means and standard deviations, as is detailed in the next section.
Afterwards, we let the process evolve and reach a stationary state.
The time necessary to reach a statistically stationary state in every simulation run is optimized by this choice of initial conditions, and is found to be less than two integral times for all simulations performed.

The above algorithm is used to build a sample path for the stochastic process in \eqref{eq:jump-scalar}. We have run this procedure for sample paths of thirty integral times in total, and three hundred sample paths were drawn for each value of $\teta$. Thus, an ensemble containing $9 \times 10^3$ integral times is built for each $\teta$, providing the significant statistics used to verify the multifractal properties of the stationary random field.

We used as parameters $\ln(\teta/T)$ ranging from $-1.0$ to $-6.0$. The more negative values correspond to more intermittency and higher Reynolds number.
The time step for the simulation was chosen to be $2 \times 10^{-3} \teta$ and the Lagrangian intermittency parameter used is $\mu = 0.3$,  which was measured in Lagrangian trajectories from direct numerical simulations in \citep{huang2014}.

Once the $X(t)$ process is calculated with this algorithm, the pseudo-dissipation $\varphi(t)$ is obtained as its exponential, from \eqref{eq:exp-x}. It was verified that the mean and standard deviation of $X(t)$ follow the analytical results \parref{eq:x2-variance} within error bars. This is particularly important for the evaluation of $\varphi$, which depends on the time periodic function $\mathbb{E}[X^2(t)]$. It is simpler and more precise to apply the analytical expression for this function \parref{eq:x2-variance} than to store the previous integral times and compute standard deviations on the fly. For our results, the first five integral times were discarded, even though the observed times to reach a stationary state were always smaller than this. These results are reported in the next section.


\section{Numerical Results} \label{sec:results}

\begin{figure}
    \centering
    \includegraphics[width=\textwidth]{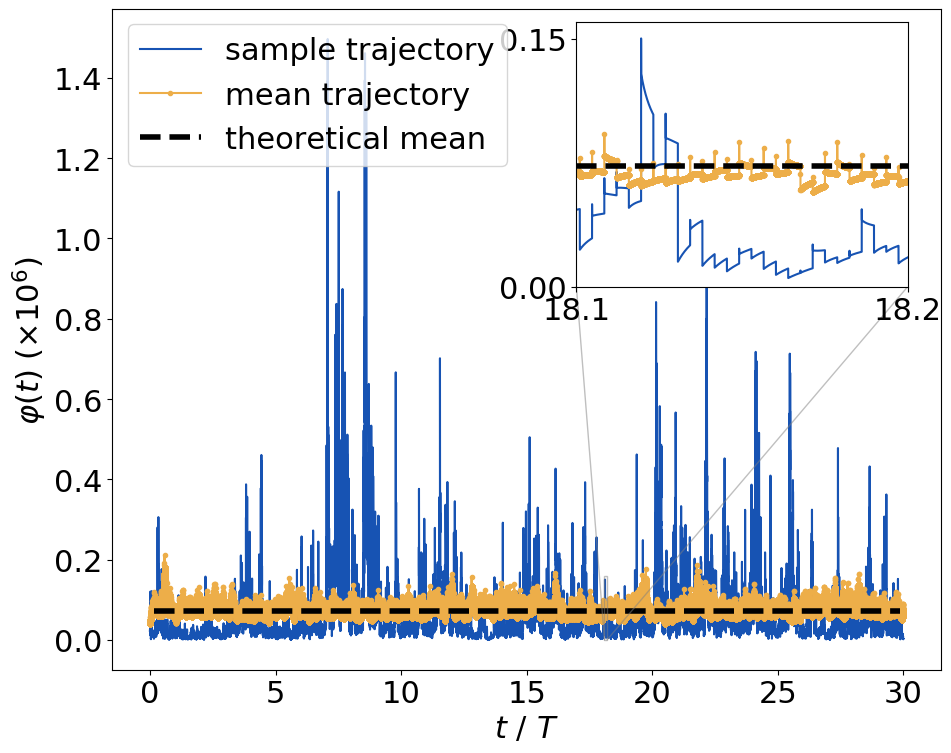}
    \caption{An illustration of the shot noise stochastic process for the energy pseudo-dissipation, $\varphi(t)$ \parref{eq:phi-x}, with multifractal properties. A sample path is drawn (blue), from an ensemble of three hundred paths, and shows strong and non-Gaussian fluctuations, characterized by localized large positive bursts. The ensemble mean (yellow) and the theoretical mean (black, dashed) are shown as well, and it can be seen that the numerical results accurately reproduce the correct average. Another noticeable feature in the ensemble trajectory is how fast the numerical solution reaches the stationary state, starting from the initial condition $\varphi(0)=1/\teta^2$.
    In this picture, $\ln(\teta/T) = -5.60$.
    The inset shows a small stretch of the full time evolution, expanded to show details of the stochastic process at small time scales, where individual jumps can be seen. The inhomogeneity of the fluctuations can also be noticed in this smaller excerpt.
    }
    \label{fig:phitraj}
\end{figure}

A sample trajectory of the shot noise multifractal process governed by Eqs.~(\ref{eq:jump-scalar}) and (\ref{eq:exp-x})
is depicted in Fig.~\ref{fig:phitraj}, along with its mean behavior.
Trajectories for this example were generated for $\ln\teta/T = -5.60$, which corresponds to one of the highest Reynolds numbers achieved in these simulations. For higher Reynolds numbers, even larger ensembles would be needed to display the same agreement between the empirical ensemble averages and theoretical predictions.
This ensemble size is sufficient for other statistical measures, though, such as probability distributions and correlations functions, because averages taken over ensembles and time translated samples can be performed.

\begin{figure}
    \centering
    \includegraphics[width=\textwidth]{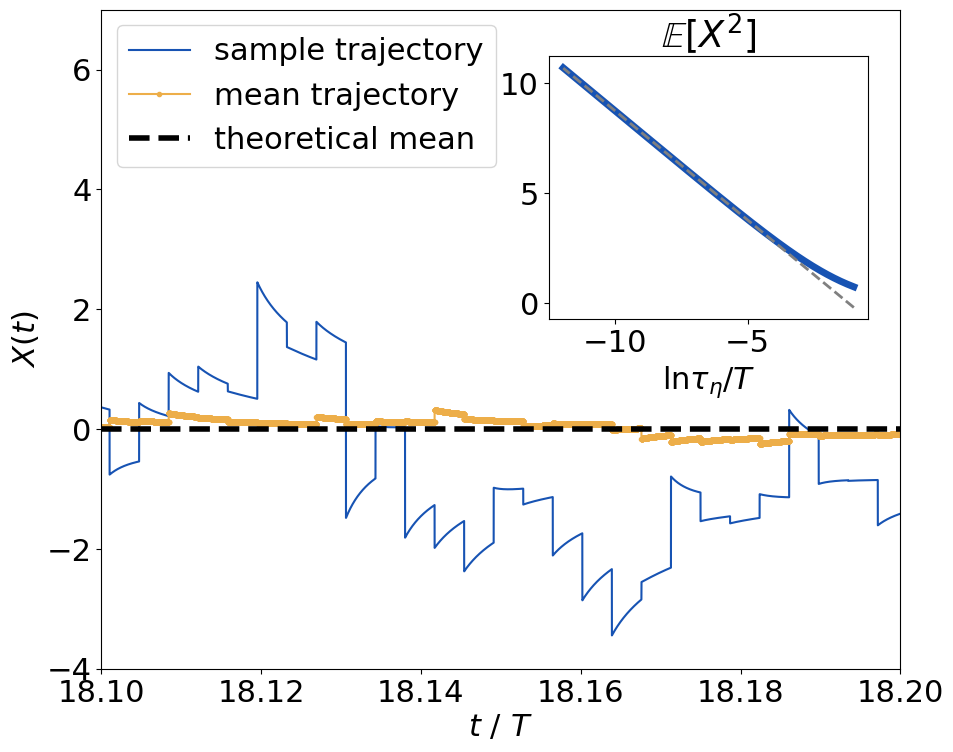}
    \caption{
    The interval depicted and the data of this figure are the same as those of the inset in Fig.~\ref{fig:phitraj}. The time evolution of the stochastic Gaussian process $X(t)$ \parref{eq:jump-scalar} is shown. The individual jumps can be seen at periodic intervals of $\teta$, and the fluctuations are much more regular, since $X$ is a Gaussian process. The colors represent the same data as in the previous figure (with $\ln(\teta/T) = -5.60$): the same sample trajectory in blue, the ensemble average in yellow and the theoretical mean in black, dashed.
    Again, the ensemble average is consistently close to the theoretical value.
    In the inset, the variance of the process $X(t)$ is shown, with its dependence in $\ln(\teta/T)$. The variance is calculated analytically with \eqref{eq:x2-variance} and a clear asymptotic linear behavior is observed as $\teta \to 0$. This linear behavior is a necessary condition for the $\varphi(t)$ field to display multifractal behavior. The gray dashed line in the inset is a linear fit in the asymptotic region to verify the linear scaling relation.}
    \label{fig:xtraj}
\end{figure}

In Fig.~\ref{fig:xtraj}, the same detailed range as the one of the inset of Fig.~\ref{fig:phitraj} is shown, which now contains the corresponding sample path of the $X(t)$ process, along with the empirical ensemble and theoretical means. The individual jumps are noticeable: They are equally likely to be positive or negative, and their intensity does not vary as vigorously as for the $\varphi(t)$ variable. The yellow curve is the ensemble average, and it is very close to the theoretical value for the mean of $X(t)$. The global character of this stochastic process is not shown, but it resembles a standard Gaussian process, since the small time scale and the periodicity of the jumps cannot be resolved if the observation window is closer to the integral scale, $T$.

In the same figure, in the inset, the asymptotic behavior of the variance of $X(t)$ is shown. For the continuous field in \citep{pereira2018}, it was demonstrated that
\begin{equation} \label{eq:lim-varx}
	\mathbb{E}\left[\left(X_P\right)^{2}\right] \stackrel[\teta \to 0]{\sim}{} \ln \left(\frac{T}{\teta}\right) \ .
\end{equation}
The equivalent relation for $X(t)$ is verified in Fig.~\ref{fig:xtraj}.
A linear fit is depicted together with the analytical curve, and the linear coefficient obtained is $0.993$. It has also been observed that this coefficient grows closer to $1.0$, the expected value for the continuous process, as the range of the fit is extended to more negative values of $\ln(\teta/T)$. This is an important property in the numerical verification that the shot noise driven process indeed displays multifractal statistics.

If we consider an instant $t$ and all other instants which differ by a multiple of $\teta$ from $t$, these points follow the discrete process described in \citep{perpete2011} for different initial conditions, and its multifractal properties can be demonstrated analytically. In particular, it is obtained that
\begin{equation} \label{eq:phi-mom-periodic}
	\mathbb{E} [ \varphi^q(t) ]_{\{t \sim t+\ell\teta\}} = \varphi_0^q \exp \left\{ \mathbb{E}[X^2(t)] K(q) \right\} \ ,
\end{equation}
in which the subscript $\{t \sim t+\ell\teta\}$ for the expectation value means that, in addition to the ensemble average, an average over all equivalent instants (separated by a multiple of the dissipative scale $\teta$) is taken as well. From this relation, taking into account \eqref{eq:lim-varx}, which has been verified numerically in Fig.~\ref{fig:xtraj}, we obtain the multifractal dependence of the statistical moments:
\begin{equation}
	\mathbb{E} [ \varphi^q(t) ]_{\{t \sim t+\ell\teta\}} = B(t) \ \varphi_0^q \left( \frac{\teta}{T} \right)^{-K(q)} ,
\end{equation}
where $B(t)$ is a function of period $\teta$.
Nevertheless, the inset of Fig.~\ref{fig:xtraj} displays the time average $\overline{\mathbb{E}[X^2(t)]}$, defined by
\begin{equation} \label{eq:x2-varmean}
	\overline{\mathbb{E}[X^2(t)]} = \frac{1}{\teta} \int_0^{\teta} \mathbb{E}[X^2(t)] \ dt \ ,
\end{equation}
reason for which there is no time dependence. Thus, the inset attests for the multifractal scaling of $\varphi$, with expectation values taken over the ensemble and time translated samples:
\begin{equation} \label{eq:moments-multifrac}
	\mathbb{E} [ \varphi^q ] = \varphi_0^q \left( \frac{\teta}{T} \right)^{-K(q)} \ .
\end{equation}

\begin{figure}
    \centering
    \includegraphics[width=\textwidth]{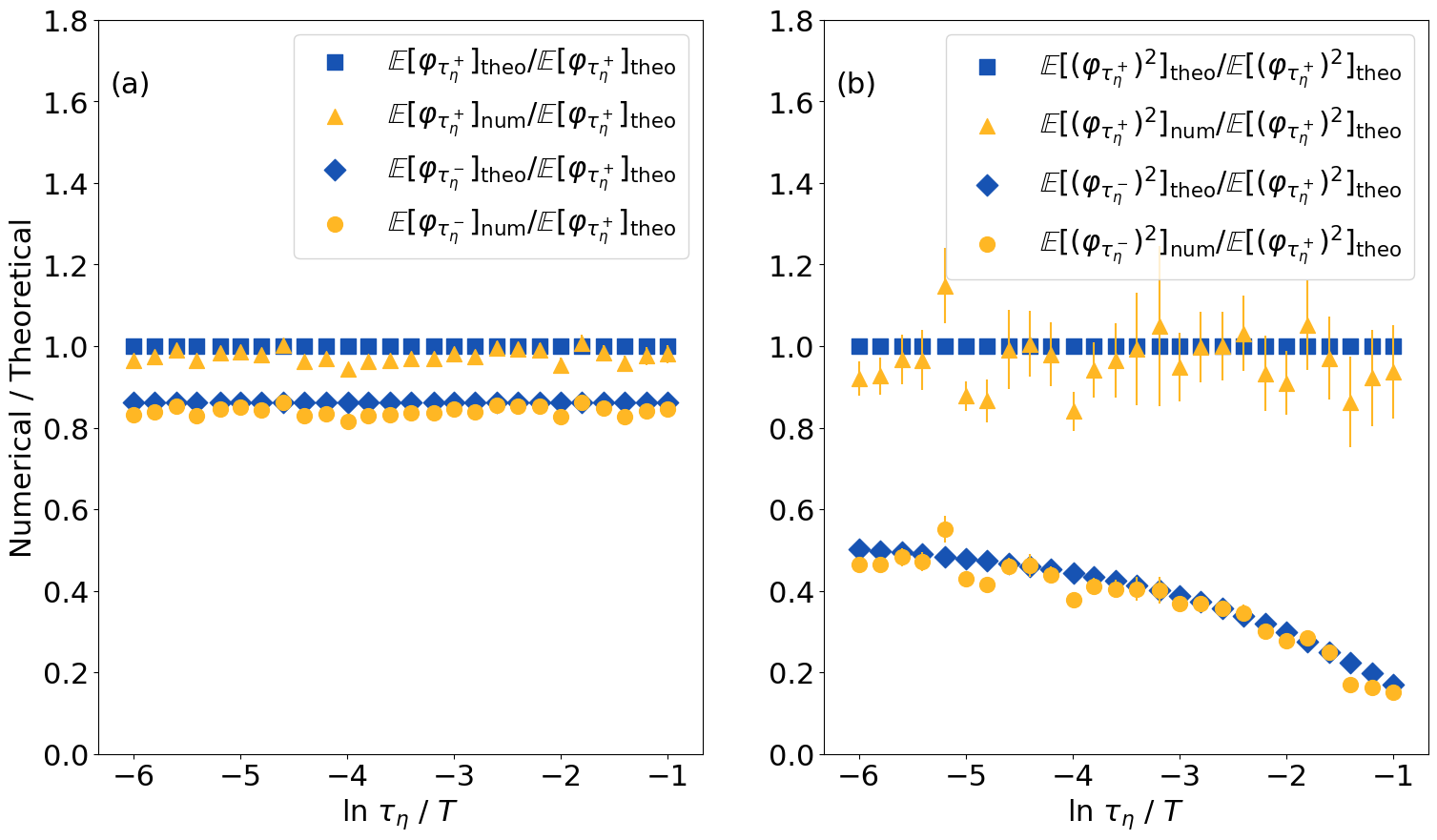}
    \caption{Comparison between low-order one-point statistical properties of the numerical solutions of \eqref{eq:jump-scalar} and their exact values. It is a consistency check on the results of the numerical calculations. The ensemble mean (a) and the variance (b) are shown. Both the mean and the variance were calculated at instants immediately before and after the jumps, and these instants are represented respectively by $\teta^-$ and $\teta^+$. Also, to make visualization more clear and the data easier to distinguish, all of the data points are given in units of the respective theoretical values after the jumps. Yellow symbols correspond to the numerical results, plotted with error bars in both cases, and blue corresponds to theoretical results.
    The values of $\ln \teta/T$ range from $-1.0$ to $-6.0$ and display all of the numerical solutions obtained.}
    \label{fig:mean_var_match}
\end{figure}

Fig.~\ref{fig:mean_var_match} is a consistency test of the numerical solution of \eqref{eq:jump-scalar}, compared with respective analytical results for the mean and variance of $\varphi(t)$ immediately before and after the jumps.
In this figure, the ensemble is larger than in the previous two figures: All independent trajectories were considered, as well as all jumps in a single trajectory. In this fashion, all points immediately before (after) a jump are equivalent in order to calculate the mean and variance of $\varphi(t)$ before (after) jumps, in the same manner as was explained in \eqref{eq:phi-mom-periodic}. The points in yellow correspond to numerical averages while those in blue correspond to theoretical values, and it can be seen that, with little exceptions, the theoretical values are within the error bars of the corresponding numerical data points. Those exceptions are expected to be corrected with a larger statistical ensemble.
The values of $\mathbb{E}[X(t)]$ and $\mathbb{E}[X^2(t)]$ vary with time in a periodic manner, and for this reason two special points in time were chosen for the analytical tests: the ones before and after the jump instants.

\begin{figure}
    \centering
    \includegraphics[width=\textwidth]{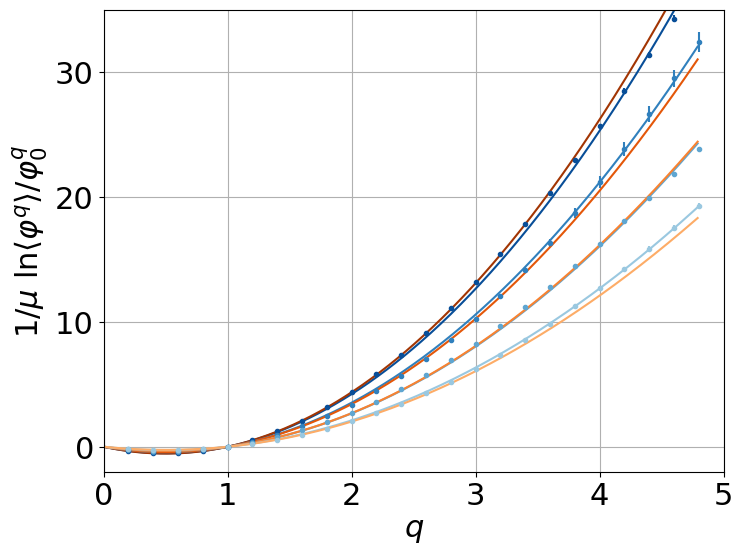}
    \caption{
    Statistical moments of the $\varphi(t)$ stochastic process, where averages are done over the ensemble and time translated samples. The numerical results correspond to the blue points, which align into a different curve for each value of $\ln \teta/T$, these curves are indicated in blue, calculated with a quadratic fit. Notice that all of the blue points include error bars. The values of $\ln\teta/T$ in this figure are $(-3.0,-3.8,-4.6,-5.6)$, with darker colors corresponding to more negative values (higher Reynolds number). In orange, theoretical curves corresponding to each of these values are displayed. These theoretical curves are quadratic, and follow the blue points and the blue curves closely for most of the calculated moments.
    These curves only deviate from each other for higher moments or higher Reynolds numbers, both regions where a significantly higher statistical ensemble would be needed.}
    \label{fig:parabola}
\end{figure}

The statistical moments $\mathbb{E}[\varphi^q(t)]$ calculated from the ensemble and time translated samples, are shown in Fig.~\ref{fig:parabola} for several values of $q$ and of $\ln \teta/T$.
This plot verifies relation (\ref{eq:moments-multifrac}), in which all time dependence has been integrated.
The numerical results, in blue points, fall in different quadratic curves according to their value of $\ln\teta/T$, in agreement with
\begin{equation}
	\mathbb{E}[\varphi^q(t)] = \varphi_0^q \exp \left\{ \overline{\mathbb{E}[X^2(t)]} K(q) \right\} \ ,
\end{equation}
with $\overline{\mathbb{E}[X^2(t)]}$ calculated from Eqs.~(\ref{eq:x2-variance}) and (\ref{eq:x2-varmean}).
This value is used to trace the orange theoretical curves in Fig.~\ref{fig:parabola}.
The data points are well approximated by parabolic fits (blue curves) which show reasonable agreement with the theoretical expectations.
Some deviation between the points and the curves are only noticeable for higher Reynolds numbers (more negative values of $\ln \teta/T$), represented by the darker curves, and for the higher moments.
The blue curves in this figure were obtained with a fit over a quadratic function $K_1(q) = a q (q-1)$, and the agreement with the points and the theoretical curves is remarkable, especially for low order moments. This result is another evidence for the lognormal behavior of the jump stochastic process.

\begin{figure}
    \centering
    \includegraphics[width=\textwidth]{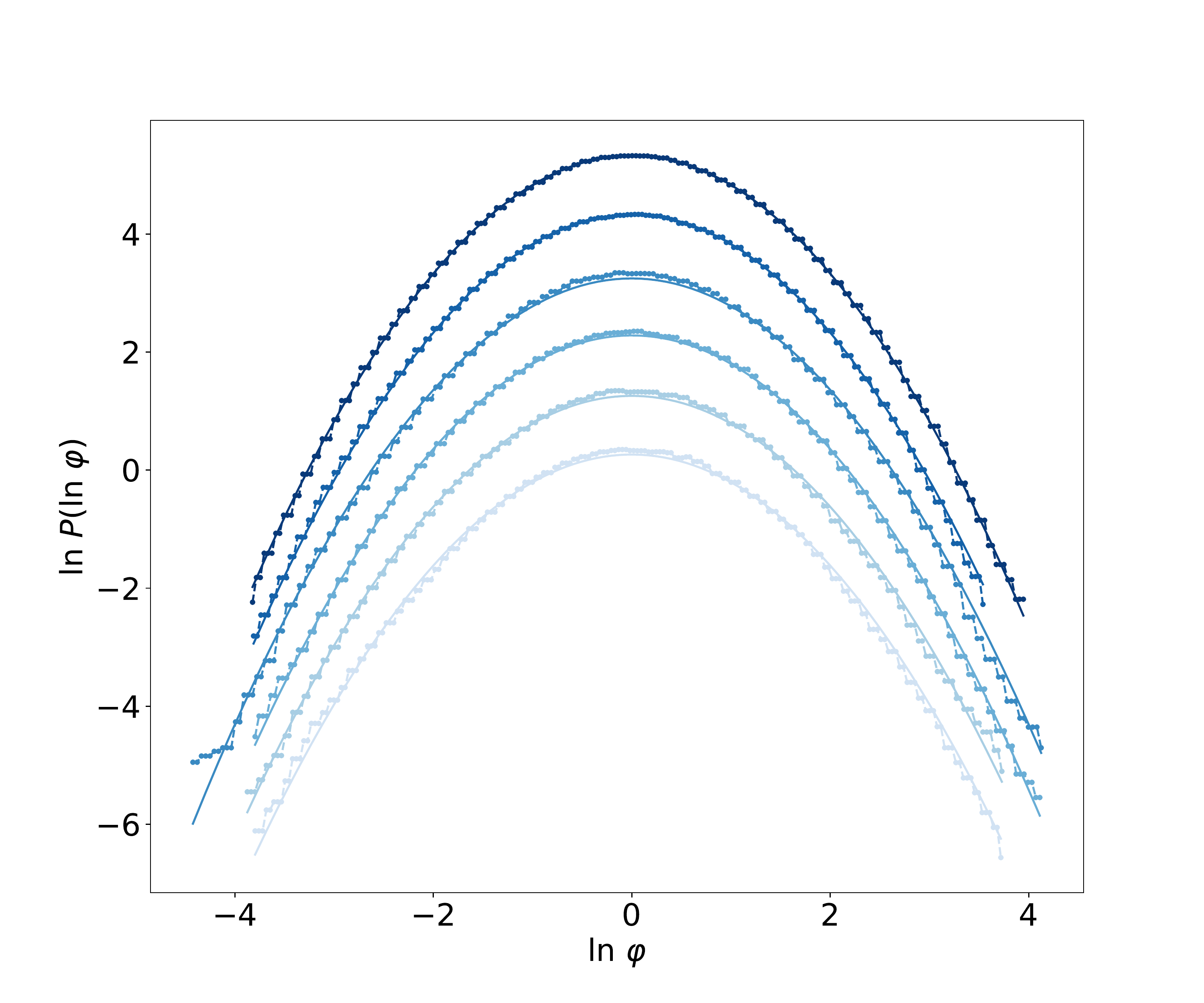}
    \caption{Normalized PDFs of $\ln \varphi$ are shown for the following values of $\ln \teta/T: (-6.0, -5.0, -4.2, -3.6, -2.8, -2.0)$, where darker colors correspond to more negative values (higher Reynolds number).
    The curves fall accurately on the continuous curves, which were obtained with a fit through a quadratic curve. This means that the probability distribution of pseudo-dissipation is lognormal for all values of $\teta$. All PDFs have been scaled to a standard Gaussian distribution (mean zero and unit variance), and they have been arbitrarily displaced upwards to simplify visualization.
    All points were obtained from the ensemble of numerical solutions of \eqref{eq:jump-scalar}, and averages over the ensembles and time translated samples have been done.}
    \label{fig:pdfs}
\end{figure}

Another form of visualizing the lognormal statistical distribution of the pseudo-dissipation $\varphi$ can be directly implemented from its probability distribution function.
They can be seen in Fig.~\ref{fig:pdfs} for several values of $\ln\teta/T$. The blue points correspond to numerically obtained PDFs, from the ensemble of numerical solutions, and the colors follow the same convention as in the other figures, with darker colors representing more negative values of $\ln\teta/T$. The mean and variance of the pseudo-dissipation have already been verified against their analytical results in Fig.~\ref{fig:mean_var_match}, hence only normalized PDFs (zero mean and unit variance) are shown in Fig.~\ref{fig:pdfs}. In this way, a direct comparison between the PDFs and an exact lognormal distribution can be done. The continuous curves are fits through quadratic functions, revealing that all of the curves fall closely on the expected normal distribution.

\begin{figure}
    \centering
    \includegraphics[width=\textwidth]{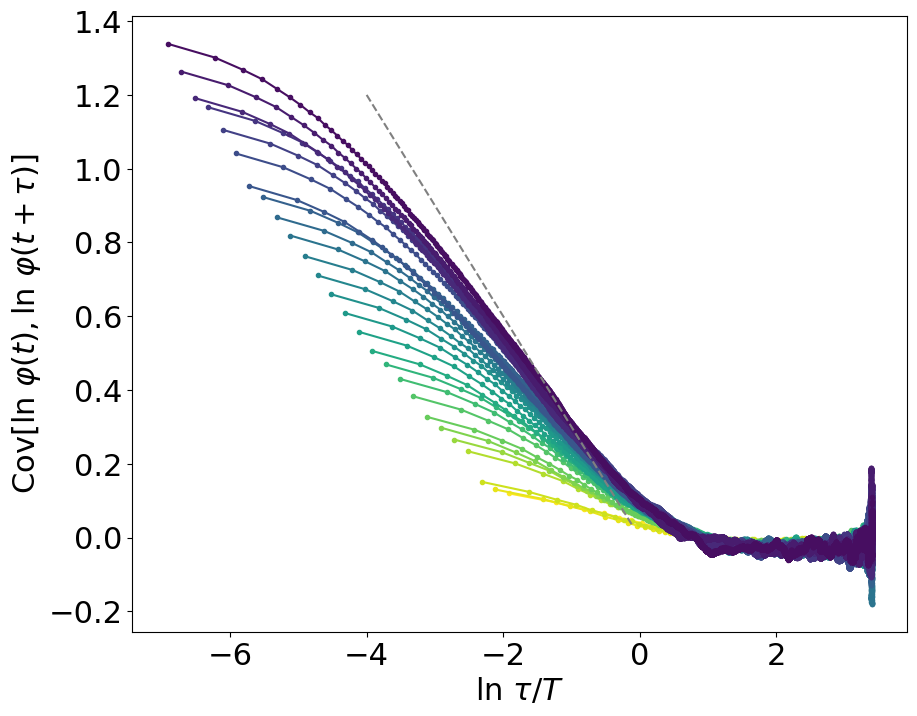}
    \caption{
    Numerical results for the autocovariance function of the pseudo-dissipation. $\tau$ is the separation between the points in this function.
    Colors range from yellow to purple, increasing in this order from less to more negative values of $\ln\teta/T$, thus the upper curves, showing a wider scaling region, are those with highest Reynolds numbers.
    The dashed line is the asymptotic relation for autocovariance in the continuous limit, where this function scales linearly with $\ln\tau/T$.
    It can be seen that as the Reynolds number grows, the region where a scaling can be seen grows, each curve becomes more closely linear, closer to the theoretical result for the continuous limit.
    }
    \label{fig:cov}
\end{figure}

Besides their lognormal behavior, another of the most relevant features of the dissipation and pseudo-dissipation statistics is their long-range correlations, which the multifractal hypothesis is able to reproduce \citep{gurvichyaglom1967,meneveausreenivasan1991,sreenivasanantonia1997}. The autocovariance of the pseudo-dissipation field, $\mathrm{Cov}[\ln \varphi(t),\ln \varphi(t+\tau)]$ has been calculated to verify the existence of long-range correlations. The autocovariance is calculated as
\begin{equation} \label{eq:gen-cov}
    \mathrm{Cov}[X,Y] =
    \mathbb{E}[(X-\langle X \rangle)(Y-\langle Y \rangle)] \ ,
\end{equation}
and the respective numerical results can be observed in Fig.~\ref{fig:cov}. In this figure, $\tau$ is the separation between two data points, where the range of interest lies in $\tau > \teta$. It can be seen that correlations grow for more negative values of $\ln\teta/T$, and as they grow, a larger scaling region can be seen for intermediate values of $\ln\tau/T$. This region is analogous to the inertial range in three-dimensional Navier-Stokes turbulence. In the scaling region, the autocovariance displays a dependence with $\ln\tau/T$ which is very close to linear, a relation which had been observed in Ref. \citep{pereira2018}. This linear dependence can be understood by rewriting the autocovariance of $\ln \varphi(t)$ as
\begin{equation}
    \mathrm{Cov}[\ln \varphi(t),\ln \varphi(t+\tau)]
    = \mu \mathbb{E}[X(t) X(t+\tau)] \ ,
\end{equation}
where a linear dependence in $\mu$ is observed. The second term, the autocorrelation of $X$, is an extension of \eqref{eq:lim-varx}, and in the limit $\teta \to 0$, it also displays a linear dependence in $\ln\tau/T$, which leads to
\begin{equation} \label{eq:phi-cov-asymp}
    \mathrm{Cov}[\ln \varphi(t),\ln \varphi(t+\tau)] \stackrel[\teta \to 0]{\sim}{} - \mu \ln \left( \frac{\tau}{T} \right) \ .
\end{equation}
The scaling region is a measure of the inertial range and is seen to grow with higher Reynolds. Also, in the gray dashed line, the exact asymptotic relation for the continuous multifractal field, \eqref{eq:phi-cov-asymp}, is shown, and it can be observed that the stochastic process with discrete jumps approaches the continuous limit as the intervals between jumps become smaller.

\begin{figure}
	\centering
    \includegraphics[width=\textwidth]{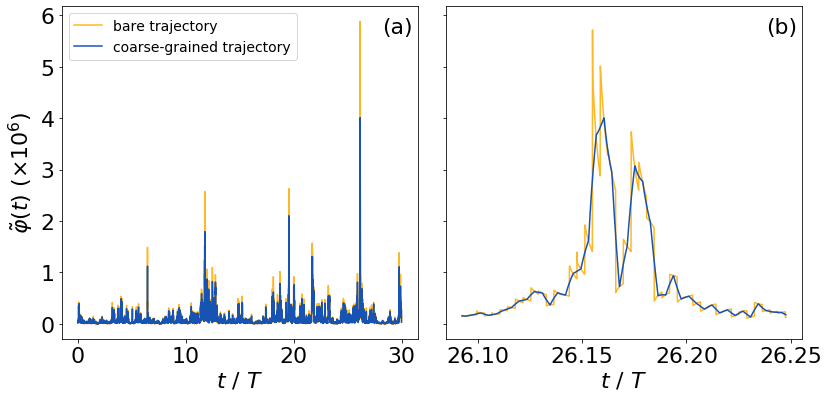}
    \caption{
    A sample trajectory of the fine-grained pseudo-dissipation field (in yellow) and its coarse-grained version, given by \eqref{eq:coarse-phi} with $\tau=\teta$ (in blue). In this figure, $\ln \teta/T = -5.60$.
    In (a), the entire time evolution can be seen, while (b) focuses on an interval around the largest fluctuation. To reliably capture high order moments, large ensembles are needed, since many intense fluctuations of the order of the one shown in this figure are needed. It is clearly observed that the coarse graining has a role in smoothing out intense fluctuations.
    }
    \label{fig:phi_smooth}
\end{figure}

\begin{figure}
	\centering
    \includegraphics[width=\textwidth]{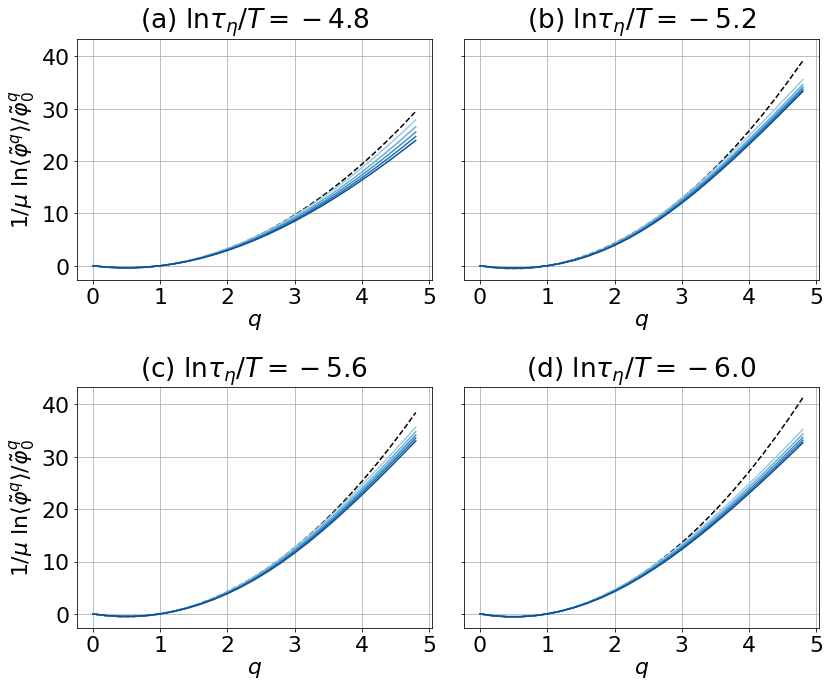}
    \caption{
    Statistical moments of the coarse-grained pseudo-dissipation field are shown for different Reynolds numbers ($\ln \teta/T = -6.0$ being the highest). Each color represents a different coarse-graining scale, from $\tau = \teta$ (lightest blue) to $\tau = 5 \teta$ (darkest blue). At lower orders, the moments at all coarse-graining scales collapse on the same quadratic curve, shown in black dashed curves, which correspond to the fine-grained moments of Fig.~\ref{fig:parabola}. At higher order, the moments deviate from the parabolic black curve, as expected for statistical moments of the coarse-grained pseudo-dissipation fields.
    }
    \label{fig:exp_smooth}
\end{figure}

\begin{figure}
    \centering
    \includegraphics[width=\textwidth]{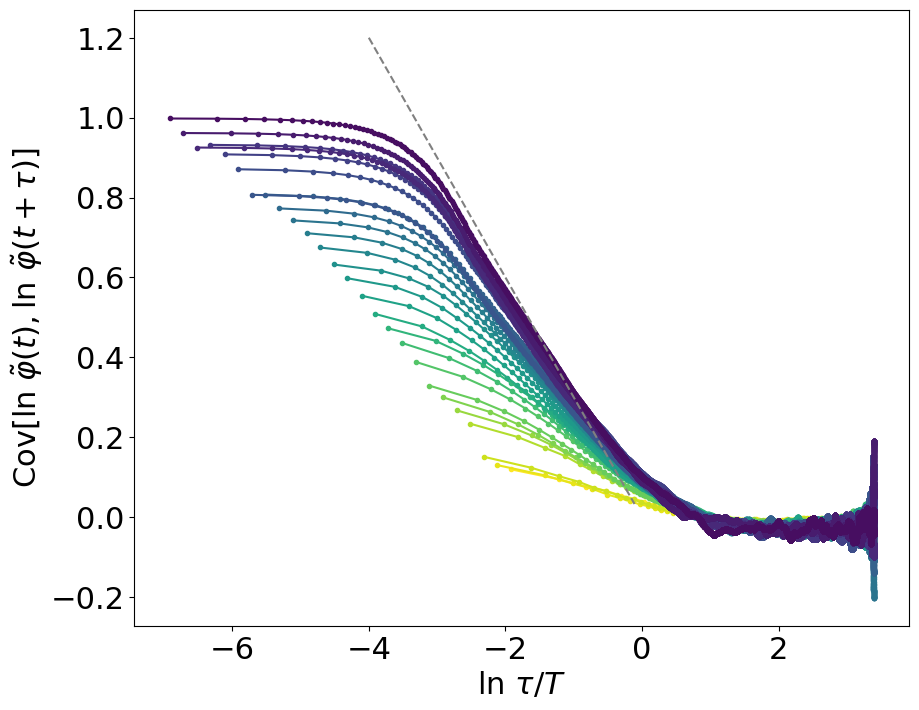}
    \caption{
    Autocovariance of the coarse-grained field $\tilde\varphi(t)$, in this figure the local averaging is done over a scale $\teta/2$.
    A clear scaling range, which is much more linear, can be seen in all of the curves, becoming more pronounced as the Reynolds number grows. Also, the slope of these linear curves is much closer to the theoretical value for the continuous limit, which is shown exactly the same as in the previous figure.}
    \label{fig:cov_smooth}
\end{figure}

These statistical properties were also investigated for time averaged fields, denoted by $\tilde \varphi(t)$ and calculated as
\begin{equation} \label{eq:coarse-phi}
	\tilde \varphi(t) = \frac{1}{\tau} \int_{t}^{t+\tau} \varphi(t') \ dt' \ ,
\end{equation}
where $\tau$ is the averaging scale under consideration.
This observable,
illustrated by
a sample trajectory in Fig.~\ref{fig:phi_smooth} with $\tau=\teta$,
is inspired by the hypothesis of refined similarity in the Lagrangian context as discussed in Section \ref{sec:stat-dissip}.
Figures are shown for the statistical moments and autocovariance of the coarse grained data, while the PDFs show no appreciable difference from their fine grained versions.
In
Fig.~\ref{fig:exp_smooth}, the exponents of the statistical moments of the coarse-grained pseudo-dissipation fields can be seen to exhibit the same quadratic behavior as the fine-grained moments. The deviations seen at higher order moments may be due to the linearization effect discussed in \citep{angeletti2011,angeletti2012}, and an investigation with larger ensembles is required to understand these differences.

For the autocovariance, which is a two point statistical observable, the behavior of the coarse-grained field is quite different from its fine-grained counterpart, yet still compatible with the asymptotic description of the continuous field.
In Fig.~\ref{fig:cov_smooth}, the autocovariance of the coarse-grained fields is seen, and the linear behavior observed in Fig.~\ref{fig:cov} for the fine-grained covariance is revealed to be even more pronounced:
The inertial range is more clearly visible, and grows as $\ln\teta/T \to -\infty$, and its slope closely approaches the theoretical value in the continuous limit.

\begin{figure}
    \centering
    \includegraphics[width=\textwidth]{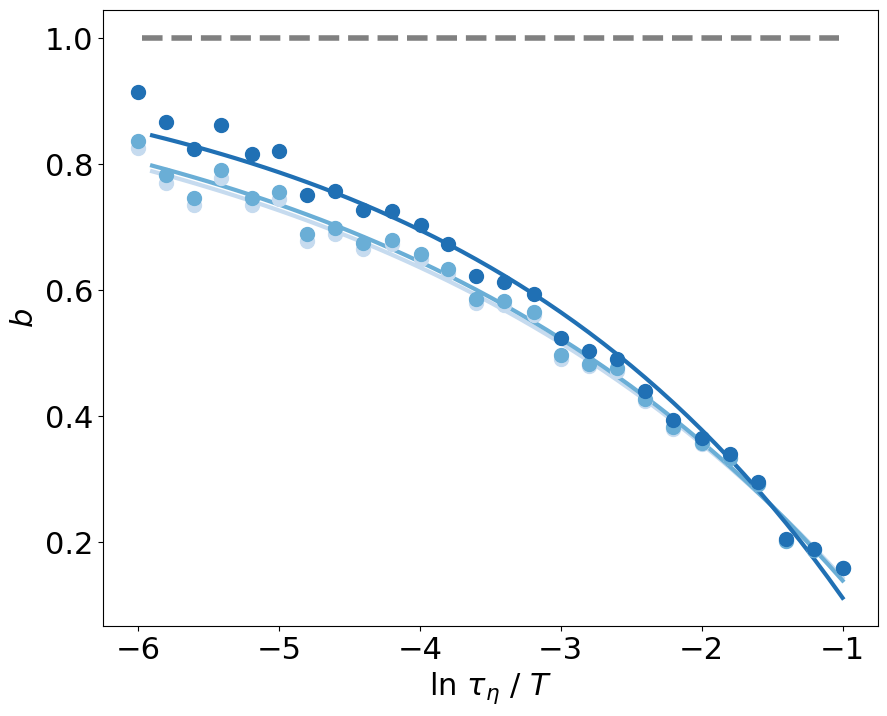}
    \caption{
    Each point has been obtained from a numerical fit of the inertial range of the autocovariance, according to \eqref{eq:fit_cov}. In this range, the asymptotic scaling \eqref{eq:fit_cov} is valid. Each color corresponds to a different coarse-graining scale, where the values shown are $\tau=(\teta/3,\teta/2,\teta)$, higher values are represented in darker colors. An exponential fit, with \eqref{eq:b_exp_fit}, through these numerical values was done to demonstrate the tendency of the data to approach the value $b=1$. This exponential fit is shown in the continuous curves. The gray dashed line on the top corresponds to $b=1$, indicating the high Reynolds number limit.}
    \label{fig:cov_irange}
\end{figure}

In order to investigate the convergence to the continuous limit, we have performed a fit of the autocovariance in the inertial range to the asymptotic functional form, that is, linear in $\ln\tau/T$, with a free parameter:
\begin{equation} \label{eq:fit_cov}
    \mathrm{Cov}[\ln \tilde\varphi(t),\ln \tilde\varphi(t+\tau)]
    = - b \mu \ln \tau/T \ .
\end{equation}
The constant $b$ is a measure of the rate of convergence to the asymptotic continuous behavior, where $b=1$. The evolution of this parameter as the dissipative scale $\teta$ changes can be seen in Fig.~\ref{fig:cov_irange}, where the points correspond to numerical fits over the respective inertial ranges. Each color represents a different coarse-graining scale $\tau$ in \eqref{eq:coarse-phi}, and as this scale grows, convergence to the continuous becomes faster. This property was observed in the autocovariance, in Fig.~\ref{fig:cov_smooth}, and is verified in Fig.~\ref{fig:cov_irange}.

Further evidence of the accelerated convergence produced by coarse-graining was obtained with a numerical fit of the curves in Fig.~\ref{fig:cov_irange}. These points slowly approach the asymptotic continuous value, $b=1$, and an exponential fit can make this argument quantitative. The function
\begin{equation} \label{eq:b_exp_fit}
    \chi(\teta) = 1 + \alpha \exp(\beta \ln \teta/T)
\end{equation}
approaches 1 as $\teta \to 0$, and is represented in the figure in continuous curves. The curves serve as a guide to the eye on the evolution of the slope $b$ as the Reynolds number grows, and furthermore show that for the higher values of $\tau$, this convergence is hastened. The exponential shape is only a plausible approximation to a curve which asymptotically approaches a value, hence fluctuations around this curve can be seen in the data. Furthermore, the inertial range is narrow for values of $\ln\teta/T$ closer to zero, which make the fit more delicate in this region. It can also be observed from Fig.~\ref{fig:cov_irange} that an increase of a few percent in the value of $b$ \parref{eq:fit_cov} would require the smallest $\teta$ to be one or two orders of magnitude lower, corresponding to a significant increase in computational effort.

\section{Conclusion} \label{sec:conclusion}

The effort to add a causal structure to the random cascade models dates back to the critiques of \citep{mandelbrot1972} to the discrete cascades.
Several approaches
have built causal stochastic processes with multifractal statistics and long-range correlations in one dimension \citep{biferale1998mimicking,boffetta1999pair,pereira2018,schmitt2003,perpete2011}.
Such random fields cannot represent Eulerian observables due to their reduced dimensionality, but they can be used to investigate the statistics of turbulence on one-dimensional Lagrangian trajectories \citep{pereira2018}.

Positive-definite quantities such as dissipation, pseudo-dissipation and enstrophy have been observed to display nearly lognormal probability distributions and long-range correlations \citep{yeungpope1989,dubrulle2019}, and such statistical properties can be understood under the multifractal formulation of turbulent flows, leading to a connection between the statistics and the geometrical properties of the energy cascade. In this work, we have built a stationary stochastic process for the pseudo-dissipation in Lagrangian trajectories, which is causal and continuous in scale ratio, and built of discrete random jumps at regular intervals.

The pseudo-dissipation random field was verified to display multifractal properties, which had already been verified for random fields driven by Wiener noise \citep{schmitt2003,pereira2018}, and for random fields defined in discrete time \citep{perpete2011}. The shot noise model for pseudo-dissipation embodies lognormal fluctuations, seen in Fig.~\ref{fig:parabola} and long-range correlations, characterized by an asymptotic logarithmic decay of its autocovariance, Fig.~\ref{fig:cov}. Despite the similarities, the present model displays regular and smooth behavior below the Kolmogorov scale, unlike the models driven by continuous noise.

We observe
that, while the jumps in this model are instantaneous and discontinuous (as seen in Figs.~\ref{fig:phitraj} and \ref{fig:xtraj}), jumps are not supposed to happen so fast in a realistic model of Lagrangian trajectories. As an effective model, the instantaneous jumps intend to capture the intermittent behavior of the pseudo-dissipation at scales larger than $\teta$, and its regular behavior at smaller scales.
Coarse-graining of the discontinuous pseudo-dissipation field might be a meaningful procedure to generate more natural fields. Even then, the procedure described here offers a skeleton to a more detailed approach to intermittent fluctuations.

Distinguishing between alternative multifractal models for the pseudo-dissipation, such as the one in this work and the continuous process in \citep{pereira2018} would require detailed measurements of high-order structure functions and of the covariance. Since the inertial range statistics of turbulence is independent on the smallest dissipative scales, it is natural to assume that large scale quantities get decoupled from small scale fluctuations. Nonetheless, the coarse-grained autocovariance in Fig.~\ref{fig:cov_smooth} exhibits considerable changes with respect to its fine-grained version. The investigation of this property in other continuous stochastic fields and in numerical data from turbulent DNS and precise experiments would produce relevant knowledge on the structure of the energy cascade and its connection to the statistical properties of Lagrangian turbulence.

The understanding of Lagrangian fluctuations is key to the effective modeling of transport properties, either of particles or fields, and to the understanding of the motion of extended structures in turbulence, such as filaments, rods and surfaces. These dynamics are heavily influenced by the localized intense bursts of energy dissipation, but still poorly understood theoretically.

These localized events,
in their turn, would certainly be affected by fluctuations of the dissipative scale \citep{paladin1987anomalous,nelkin1990multifractal,frischvergassola1991,yakhot2006probability,arneodo2008universal}, which is considered fixed at $\teta$ in this model. More accurate multifractal stochastic models would have to address the connection between sub-Kolmogorov statistics in Lagrangian trajectories and the energy cascade, a connection which is fundamental for the correct statistics of coarse-grained observables, as noticed in Ref.~\citep{arneodo2008universal}.
While being an essentially open modeling problem, an alternative and hopefully more realistic model could, in 
principle, be devised in terms of smoother noise jumps where its bandwidth is correlated with fluctuations of the pseudo-dissipation field, in order to account for the fluctuations of the dissipative scale and corrections to the lognormal statistics as well.

Another extension of this approach is the understanding of the full spatio-temporal structure of fluctuations in Eulerian turbulence in terms of a causal stochastic process. This would provide a way to establish a connection between the Navier-Stokes equations and the stochastic models for the energy cascade, a huge step in the understanding of the Richardson cascade and the geometrical properties of turbulence, and on the origin of its multifractal statistical features \citep{lejan1997,cardesa2017,doan2018,motoori2019}.

\section{Acknowledgments}

G.B.A. thanks CNPq for financial support. L.M. thanks CNPq and Petrobras (COPPETEC 20459) for partial support.
The authors thank R. M. Pereira, L. Chevillard, F. Ramos
and D. Rodrigues for helpful discussions,
and NIDF (N\'{u}cleo Interdisciplinar de Din\^{a}mica de Fluidos) for
the use of its facilities and computational resources.

\appendix
\section{It\^{o}'s Lemma for Pure Jump Processes}

As was done in Refs.~\citep{schmitt2003,pereira2018}, a dynamical equation for the pseudo-dissipation itself can be obtained from the dynamical equation for $X(t)$, \eqref{eq:jump-scalar}, and the relation between the $X$ and $\varphi$ variables, \eqref{eq:exp-x}.
Consider for a moment the general stochastic differential equation
\begin{equation} \label{eq:x-sde}
	dX(t) = F(t^-, X(t^-)) \ dt + \sum_{0 \leq t_{\ell} \leq t} G(t^-,X(t^-)) \delta(t-t_{\ell}) \alpha_{\ell} \ dt \ ,
\end{equation}
where $F$ and $G$ are arbitrary functions of $t$ and $X(t)$, respectively called the drift and jump terms. This equation does not have any continuous noise term (proportional to a Wiener measure $dW(t)$), because the stochastic differential equation proposed in this work does not possess the Wiener term either.
In addition, an appropriate set of initial conditions for $X(t)$ is provided.
The new variable, $Y$, is obtained from the original variable through an arbitrary continuous function $f$, as
\begin{equation} \label{eq:f}
	Y(t) = f(t,X(t)) \ .
\end{equation}
A stochastic differential equation for $Y(t)$ is obtained with It\^{o}'s lemma for semimartingales, which is the appropriate expression for a change of variables in a stochastic process, equivalent to the chain rule in standard calculus \citep{protter2005,klebaner2012}. Semimartingales are generalizations of local martingales: While the latter are represented by continuous stochastic processes, such as the standard Brownian motion, the former may display discontinuous jumps, which are central to the current discussion. The solution $X(t)$ of \eqref{eq:x-sde} is thus a semimartingale.

In its semimartingale formulation, It\^{o}'s lemma is expressed as
\begin{equation} \label{eq:ito-original}
    \eqalign{
    &Y(t) = Y(0) +
    \int_0^t \partial f(s^-,X(s^-)) / \partial s \ ds \\
    &+ \int_0^t f'(s^-,X(s^-)) dX(s) + \frac12 \int_0^t f''(s^-,X(s^-)) d[X,X]^c(s) \\
    &+ \sum_{0 \leq t_{\ell} \leq t} \Big( f(t_{\ell},X(t_{\ell})) - f(t_{\ell}^-,X(t_{\ell}^-))
    - f'(X(t_{\ell}^-)) (X(t_{\ell}) - X(t_{\ell}^-)) \Big) \ .}
\end{equation}
The integration interval, from $0$ to $t$, includes several jump instants, denoted by $t_{\ell}$ with an integer index $\ell$ differentiating each jump. Because of the discontinuities, it is important to prescribe that the $X(t)$ process is \textit{c\`{a}dl\`{a}g}, which means that terms of the form $X(s^-)$ should be calculated as the limit
\begin{equation}
	X(s^-) = \lim_{t \to s^-} X(t) \ .
\end{equation}
If $s$ is a jump instant, this limit does not include the contribution from the discontinuous jump, which is only accounted for in $X(s)$. Whereas if $s$ is not a jump instant, $X(s)$ and $f(s,X(s))$ are continuous at this point.
The first four terms in the RHS of \eqref{eq:ito-original} are exactly equal to those in It\^{o}'s lemma for continuous processes, with the only difference that the discontinuous jumps require a distinction between left and right limits. As in the traditional It\^{o}'s lemma, the derivatives $f'(t,X(t))$ and $f''(t,X(t))$ are taken with respect to $X(t)$.

The \textit{continuous quadratic variation} $[X,X]^c(t)$ of the Wiener process is simply $t$, concluding the identification with the lemma for local martingales.
In general, the quadratic variation is defined by
\begin{equation} \label{eq:quad-var}
	[X,X]_{t}=\lim _{\delta t \rightarrow 0} \sum_{k=1}^{n}\left(X_{t_{k}}-X_{t_{k-1}}\right)^{2} \ ,
\end{equation}
where time has been partitioned into $n$ intervals of size $\delta t_k = t_k - t_{k-1}$ and $\delta t$ is the maximum size among these partitions \citep{protter2005,klebaner2012}. The continuous quadratic variation is the continuous part of \eqref{eq:quad-var}.
If the stochastic force is purely jump-discontinuous, as is the case in \eqref{eq:jump-scalar}, its continuous quadratic variation is zero. Also, using \eqref{eq:x-sde}, we notice that the discontinuity $X(t_{\ell}) - X(t_{\ell}^-)$ which appears in \eqref{eq:ito-original} is equal to $G(t_{\ell}^-,X(t_{\ell}^-)) \alpha_{\ell}$.
Thus, replacing \eqref{eq:x-sde} in \eqref{eq:ito-original}, one of the terms in $f'(s^-,X(s^-)) dX(s)$ is canceled by $f'(t_{\ell}^-,X(t_{\ell}^-)) (X(t_{\ell}) - X(t_{\ell}^-))$. With this, we obtain \textit{It\^{o}'s lemma for pure jump processes}:
\begin{equation} \label{eq:ito-semi}
\eqalign{
    Y(t) &= Y(0) +
    \int_0^t \frac{\partial f}{\partial s}(s^-) \ ds
    + \int_0^t f'(X(s^-)) F(s^-,X(s^-)) ds \\
    &+ \sum_{\ell} \Big( f(t_{\ell},X(t_{\ell})) - f(t_{\ell}^-,X(t_{\ell}^-)) \Big) \ .
    }
\end{equation}
In differential notation, this is equivalent to
\begin{equation} \label{eq:ito-jump}
\eqalign{
    d Y(t) &=
    \partial f / \partial t \ dt +
    f'(X(t^-)) F(t^-,X(t^-)) dt \\
    &+ \sum_{\ell} \Big( f(t_{\ell},X(t_{\ell})) - f(t_{\ell}^-,X(t_{\ell}^-)) \Big)
    \delta(t-t_{\ell}) dt \ . }
\end{equation}

At first glance, this definition may look circular, because the variable $Y$ and the variable $X$ appear simultaneously. In fact, only the initial condition for $X(t)$ is needed, which is easily converted to an initial condition for $Y(t)$. All other appearances of $X(t)$ in \eqref{eq:ito-jump} are causal, referring to values of $Y(t)$ already calculated, thus $X(t) = f^{-1}(t,Y(t))$. The term $f(t_{\ell},X(t_{\ell}))$, when $t_{\ell}$ is a jump instant, needs the value of $X$ at the current instant, which is simply the left-limit at $t_{\ell}^-$ with the random contribution added:
\begin{equation} \label{eq:x-jump-gen}
	X(t_{\ell}) = X(t_{\ell}^-) + G(t_{\ell}^-,X(t_{\ell}^-)) \alpha_{\ell} \ .
\end{equation}
Thus, \eqref{eq:ito-jump} is an entirely self-consistent way to determine the time evolution of the random field $Y(t)$.

In the specific model considered in this work,
$X(t)$ is a stochastic process with Gaussian fluctuations and its exponential is the variable of interest, with lognormal fluctuations and long-range correlations.
Through It\^{o}'s lemma \parref{eq:ito-semi}, we obtain a stochastic differential equation for the pseudo-dissipation field, $\varphi(t) = f(X(t))$, defined in \eqref{eq:exp-x}.

The equation we obtain through this procedure for the pseudo-dissipation field is
\begin{equation} \label{eq:phi-x}
\eqalign{
    &d\varphi(t) = \varphi(t^-)
    \Bigg( - \frac{1}{T} \ln \frac{\varphi(t^-)}{\varphi(0)} -     \frac{\mu}{2 T} \mathbb{E}[X^2(t^-)] + \sqrt{\mu} \beta(t) \\ &- \frac{\mu}{2} \frac{\partial \mathbb{E}[X^2(t^-)]/2 }{\partial t} \Bigg) \ dt + \sum_{\ell} \Big( f(\varphi(\teta \ell)) - f(\varphi(\teta \ell^-)) \Big) \delta(t-\teta \ell) dt \ .
		}
\end{equation}
For this process to be completely well defined, we only need an initial condition for the field $\varphi$ (or equivalently for $X$). Since this stochastic process has a long-term memory, it is necessary to provide $X(s)$ for $s \in ]-T,0]$, corresponding to the past time-evolution of $X$. After a few integral times, the influence of the initial condition vanishes, and the process reaches a stationary state.


\newcommand{\newblock}{}
\bibliographystyle{unsrtnat}
\bibliography{bib}

\end{document}